\documentclass[reprint,aps,prb,twocolumn,floats,showpacs,footinbib,superscriptaddress]{revtex4-1}

\usepackage{bm} 
\usepackage{graphicx}
\usepackage{amsmath}
\usepackage{amsfonts}
\usepackage{amssymb}
\usepackage{bbold}
\usepackage{times} 
\usepackage{latexsym}

\newcommand{\ua}{\uparrow}
\newcommand{\da}{\downarrow}

\newcommand{\keta}{ \rangle}
\newcommand{\bra}[1]{\left \langle {#1} \right|}

\newcommand{\ket}[1]{\left| {#1}\right \rangle}

\newcommand{\sfr}[1]{ \frac{1}{#1} }

\newcommand{\mubg}{\mu_B g_0}
\newcommand{\mub}{\mu_B}
\newcommand{\phii}{\varphi}
\newcommand{\kp}{\kappa}
\newcommand{\ii}{\mathbf{i}}

\newcommand{\veps}{\varepsilon}

\newcommand{\vecb}[1]{\mathbf{#1}}

\newcommand{\beq}{ \begin{equation} }
\newcommand{\eeq}{\end{equation}}

\usepackage[usenames]{color}

\def \PAST{Institute of Experimental Physics, Faculty of Physics,
University of Warsaw, Pasteura 5, 02-093 Warsaw, Poland}
\def \IFPAN{Institute of Physics, Polish Academy of Sciences, Aleja Lotnik\'ow 32/36, 02-668 Warsaw, Poland}

\begin{document} 

\title{Anisotropy of in-plane hole g-factor in CdTe/ZnTe quantum dots} 
\author{A. \surname{Bogucki}}\email{Aleksander.Bogucki@fuw.edu.pl}\affiliation{\PAST}
\author{T. \surname{Smole\'nski}}\email{Tomasz.Smolenski@fuw.edu.pl}\affiliation{\PAST}
\author{M. \surname{Goryca}}\affiliation{\PAST}
\author{T. \surname{Kazimierczuk}}\affiliation{\PAST}
\author{J. \surname{Kobak}}\affiliation{\PAST} 
\author{W. \surname{Pacuski}}\affiliation{\PAST}
\author{P. \surname{Wojnar}}\affiliation{\IFPAN}  
\author{P. \surname{Kossacki}}\affiliation{\PAST} 

\date{\today}

\begin{abstract} 

Optical studies of a bright exciton provide only limited information about the hole anisotropy in a quantum dot. In this work we present a universal method to study heavy hole anisotropy using a dark exciton in a moderate in-plane magnetic field. By analysis of the linear polarization of the dark exciton photoluminescence we identify both isotropic and anisotropic contributions to the hole g-factor. We employ this method for a number of individual self-assembled CdTe/ZnTe quantum dots, demonstrating a variety of behaviors of in-plane hole g-factor: from almost fully anisotropic to almost isotropic. We conclude that, in general, both contributions play an important role and neither contribution can be neglected. 
\end{abstract}

\pacs{75.75.-c; 78.67.Hc; 78.55.Et}

\maketitle

\section{Introduction}
A solid state system, e.g. a semiconductor quantum dot (QD), can be studied by subjecting it to various perturbations: electric \cite{Bennett_2013_NC,Ares_2013_PRL} or magnetic field \cite{Maksym_1990_PRL,Bayer_1999_PRL}, axial or hydrostatic strain \cite{Nakaoka_2005_PRB,Bryant_2011_PRB,Yeo_2013_NN}, shape or composition variation \cite{Kumar_2006_Small,Shumway_2001_PRB,Schlereth_2008_N}, photonic environment\cite{Gerard_1998_PRL,Jakubczyk_2014_ACSNN}, etc. Out of these possibilities, a magnetic field stands out as a very universal one, since its application does not rely on specific structure of the sample.  Indeed, magnetic properties of QDs have been extensively studied in a number of different material systems. In the first order, the magnetic field modifies the energy of the excitons in a QD due to the Zeeman effect. The strength of this effect is determined by the g-factors of the confined carriers. Due to the band structure of zinc-blende or wurzite semiconductors\cite{Chuang__1995}, the electron g-factor is typically considered isotropic. Conversely, in an idealized case the hole ground state in the epitaxial QD has pure heavy-hole character \cite{Bayer_1999_PRL,Bayer_2002_PRB} and thus its g-factor is fully anisotropic ($g_x = g_y = 0$, $g_z \neq 0$). 

In the real QD structures, the hole has typically non-zero in-plane
g-factor\cite{Koudinov_2004_PRB,Kowalik_2007_PRB,Leger_2007_PRB,Witek_2011_PRB,Schwan_2011_APL}. This in-plane g-factor was identified to originate from two distinct effects\cite{Koudinov_2004_PRB,Kowalik_2007_PRB}: valence band mixing and the cubic term of Luttinger-Kohn Hamiltonian. These two contributions differ in $x-y$ anisotropy of the resulting in-plane hole g-factor. This issue has been studied so far mostly in the context of reduction of the fine structure splitting between two bright excitons in a QD\cite{Stevenson_2006_PRB,Kowalik_2007_PRB,Pooley_2014_PRA,Mrowinski_APL_2015}. 

In our work, we introduce an efficient method of analyzing the impact of these two mechanisms using a dark exciton in a QD. The dark excitons, i.e., the excitons with parallel orientation of electron and hole spins, are characterized by a relatively long lifetime\cite{Smolenski_2012_PRB,Smolenski_2015_PRB} and a small zero-field energy splitting\cite{Bayer_2002_PRB,Schwartz_2015_PRX}. Such properties makes the dark exciton a reasonable candidate for quantum computing. However, due to its weak coupling to photons, the dark-exciton-based qubit was demonstrated only recently\cite{Schwartz_2015_PRX}. The coupling of the dark exciton to photons can be increased by application of the in-plane magnetic field\cite{Bayer_2000_PRB,Koudinov_2004_PRB,Kowalik_2007_PRB, Leger_2007_PRB,Smolenski_2012_PRB}. In such a case, a dark exciton state gains an admixture of a bright exciton, which increases its oscillator strength. We show that under such conditions the polarization of the dark exciton luminescence is a sensitive probe of the in-plane hole g-factor. The main advantage of using the dark exciton is related to its zero-field splitting.
The bright exciton states are subjected to substantial anisotropic fine structure splitting, which determines their polarization properties. Conversely, the fine structure splitting of the dark exciton is usually more than one order of magnitude smaller and therefore it does not hinder the subtle polarization effects related to light-hole heavy-hole mixing. The angular field dependence studies of dark exciton polarization give an insight into the relative contribution of the two possible mechanisms to the in-plane hole g-factor. Our results demonstrate that neither of these two possible mechanisms can be neglected as they contribute to the hole g-factor to a similar extent.

\section{Linear polarization of the dark exciton}

\subsection{Samples and experimental setup}
In the experiment we used two types of samples grown by molecular beam epitaxy (MBE). Both types of samples contain self-assembled CdTe QDs embedded in ZnTe barrier formed during the evaporation of amorphous tellurium layer as proposed by Tinjod \textit{et al.} \cite{Tinjod_2003_APL}. The main difference between these types is the QD formation temperature. The first type (LT) consists of a sample with low formation temperature $T=260^\circ$~C. In case of the second type (HT), the QDs were formed at higher temperature $T=345^\circ$~C. The growth procedures used for LT samples were similar to methods presented in Refs. \onlinecite{Wojnar_2008_N} and \onlinecite{Wojnar_JCG_2011}. More detailed description of HT samples growth procedures can be found in Refs. \onlinecite{Kobak_2013_JCG,Kobak_2014_NC}. The only difference between QDs presented in above references and our samples is the fact that in our QDs no magnetic ions were present.

For optical experiments, the sample was placed in liquid helium bath cryostat ($T=1.5$~K). The cryostat was equipped with two pairs of superconducting split coils. By controlling the currents of both coils we were able to apply in-plane magnetic field of up to 2~T in any direction.

The samples were excited with a 405~nm continuous-wave (CW) semiconductor laser focused by immersive mirror objective to a spot of diameter smaller than 1~$\mu$m. Photoluminescence (PL) spectra were recorded using 0.75~m spectrograph and a CCD camera. Repetitive measurements of a PL polarization dependence were carried out by a rotating motorized $\lambda/2$ plate in front of a linear polarizer in the detection path.

\subsection{Experimental results}

\begin{figure}[h]
\includegraphics[width=86mm]{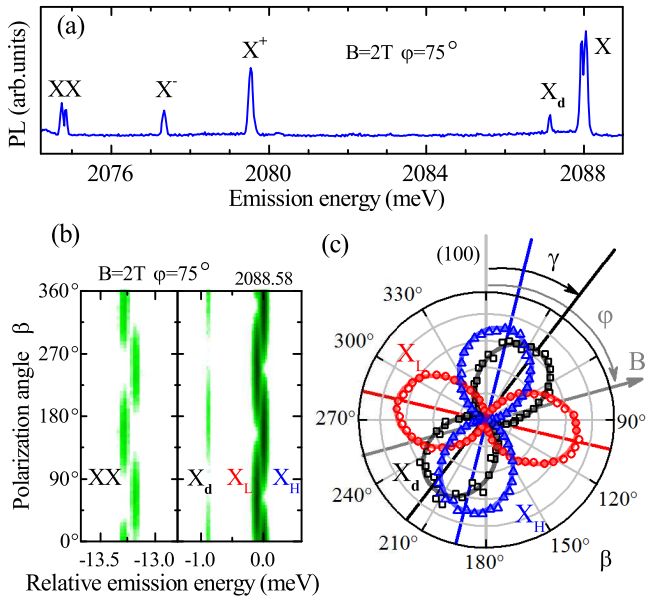}
\caption{(Color online) (a) A typical PL spectrum of a single self-assembled CdTe/ZnTe QD under in-plane magnetic field $B=2$~T (Voigt configuration). (b) False-color map presenting PL spectra of the same QD for different directions of detected linear polarization and (c) corresponding polar plots of X$_\mathrm{d}$ and X emission lines intensities vs. detection polarization angle $\beta$.}
\label{fig:spectrum_butterfly_75perc}
\end{figure}

Figure \ref{fig:spectrum_butterfly_75perc}(a) presents a typical PL spectrum
of a self-assembled CdTe/ZnTe QD in a transverse magnetic field. The spectrum consists of a series of lines corresponding to recombination of different excitonic complexes\cite{Brunner_1994_PRL,Besombes_2002_PRB,Kazimierczuk_2011_PRB}. The applied magnetic field of 2~T is weak enough to neglect the Zeeman shift of the studied transitions, but sufficient to make the dark exciton line visible.

In this work we focus on a~neutral exciton. Its main PL lines denoted in Fig. \ref{fig:spectrum_butterfly_75perc}(a) as X are related to the bright exciton states with antiparallel spins of the electron and the hole. The energy splitting $\delta_1$ between the two bright exciton lines originates from the anisotropic electron-hole exchange interaction and corresponds to about $0.2$~meV in case of CdTe/ZnTe QDs\cite{Kazimierczuk_2011_PRB}. The orientation of this anisotropy can be easily determined by measuring the linear polarization of the bright exciton lines, as shown in Fig. \ref{fig:spectrum_butterfly_75perc}(b). The same measurement serves also as a confirmation of the correct identification of the neutral exciton \cite{Bayer_2002_PRB,Suffczynski_2006_PRB,Suffczynski_2008_PRB}.

Due to the presence of transverse magnetic field, the PL spectrum of a QD also features a line related to the radiative recombination of a dark exciton X$_d$. It is observed approximately $1$~meV below the bright exciton lines. This distance corresponds to the energy of isotropic electron-hole exchange interaction $\delta_0$.
 
As seen in Figs. \ref{fig:spectrum_butterfly_75perc}(b) and \ref{fig:spectrum_butterfly_75perc}(c), the X$_d$ line is linearly polarized. Importantly, the direction of its linear polarization, in general, does not coincide with any other key direction: neither bright exciton anisotropy axis, (100) crystallographic axis, nor the direction of the magnetic field \cite{Tonin_2012_PRB}. The angle of polarization direction of the dark exciton emission line will be further on referred to as $\gamma$. 

As we show in the following section, the orientation of X$_d$ polarization depends on the magnetic field orientation in a complex manner due to the tensor character of the hole g-factor. Experimentally, we access this dependence by measuring the dark exciton PL intensity for different orientations $\beta$ of detected linear polarization and different orientations $\phii$ of the in-plane magnetic field. The angles $\beta$, $\phii$ and $\gamma$ are measured in the laboratory frame (Fig. \ref{fig:spectrum_butterfly_75perc}(c)), in which the crystallographic direction (100) is vertical.

\begin{figure}[b]
\begin{center}
\includegraphics[width=86mm]{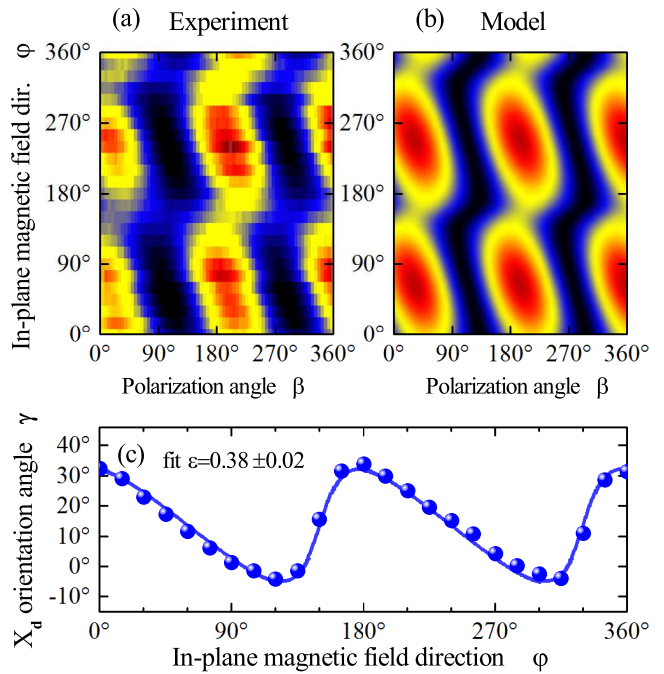}
\end{center}
\caption[]{(Color online)  (a) Measured intensity map of brightened dark exciton line for different detection polarization angles and different directions of in-plane magnetic field. The color represents the intensity of the dark exciton line. (b) Simulation of such a map based on the model described in section \ref{sec:model}. The procedure used for determination of light polarization was the same as in Refs. \onlinecite{Poem_PRB_2007,Kazimierczuk_2011_PRB}   (c) Brightened dark exciton emission line polarization angle $\gamma$ for different directions of in-plane magnetic field $\phii$. Symbols represent measured data, while the solid line represents the fitted curve described by Eq. (\ref{eq:gamma}).
}
\label{fig:example_map_and_fit}
\end{figure}

Typical results of the described measurement are presented in Fig. \ref{fig:example_map_and_fit}(a).  As it is expected for an almost fully polarized dark exciton line (see Appendix \ref{app:matrix}), the emission intensity changes like $\cos ^2 \beta$. However, for different orientations of the magnetic field $\phii$ the direction corresponding to maximum PL intensity (i.e., angle $\gamma$) varies. In order to present this effect in a concise form, we fit the data for each field direction by $I=A \cos^2(\beta-\gamma)+C$. The example plot of the extracted $\gamma$ as a function of $\phii$ is shown in Fig. \ref{fig:example_map_and_fit}(c). 


\subsection{Theoretical model}
\label{sec:model}

The usability of brightened dark exciton for determining anisotropy properties of the hole g-factor tensor stems from the strong dependence of its optical properties on the magnetic field perpendicular to the growth axis. The dark exciton, which has total angular momentum of $2$, is optically inactive, as far as the hole ground state is made of pure heavy-hole (HH) state. However, real self-assembled QDs exhibit shape imperfections and feature non-trivial strain distribution, which results in valence band mixing. In the first approximation, the two lowest-energy hole states might be expressed as
\begin{equation}
\label{eq:mixing}
\ket{\phi^{\pm}_{H}}= \left(\ket{\pm 3/2}+\lambda e^{\pm \ii 2 \theta} \ket{\mp 1/2}\right)/\sqrt{1+\lambda^2}
\end{equation}
where $\lambda$ describes the strength of the valence band mixing, while $\theta$ is an effective hole anisotropy direction. It should be stressed that this direction is not the same as the orientation of electron-hole exchange interaction anisotropy. In the presence of a light-hole (LH) admixture in the hole ground state the dark exciton becomes optically active and can emit photons in the direction perpendicular to the QD growth axis\cite{Karlsson_PRB_2010,Dupertuis_2011_PRL,Smolenski_2012_PRB}. In order to enable the dark exciton emission along the QD growth axis we apply a transverse magnetic field, which results in mixing of bright and dark excitonic states\cite{Bayer_2000_PRB,Koudinov_2004_PRB,Kowalik_2007_PRB,Leger_2007_PRB,Smolenski_2012_PRB}. We note that some additional brightening of the dark exciton may arise due to the QD symmetry reduction \cite{Schwartz_2015_PRX,Zielinski_2015_PRB}. However, in our experiments we do not observe any signature of such an effect and thus neglect it in further considerations.

The behavior of the exciton in the magnetic field is governed by the g-factors of the constituting carriers. Electron in magnetic field is described by the Zeeman Hamiltonian $\hat{H}^e_B=\mub \vecb{S} \hat{g}_e \vecb{B}$, where $\mub$ is the Bohr magneton, $\vecb{S}$ is 1/2 spin operator, $\vecb{B}$ is the magnetic field vector and $\hat{g}_e$ is the electron g-factor tensor. In the usual case, the electron g-factor is isotropic\cite{Schwan_2011_APL}, thus we consider $\hat{g}_e$ to be a scalar.

The hole in magnetic field is described by the Luttinger-Kohn Hamiltonian (LK) \cite{Luttinger_1956_PR}
\begin{equation}
 \hat{H}^h_B=\mubg[\kp\vecb{J}\vecb{B}+q(J^3_xB_x+J^3_y B_y+J^3_z
B_z)], \label{eq:luttinger_hamiltonian}
\end{equation}
where $g_0$ is a free electron g-factor, $\vecb{J}$ is 3/2 momentum operator and $q,\,\kp$ are Luttinger parameters. The first term in the above Hamiltonian will be further called Zeeman part of the LK Hamiltonian because of its similarity to standard Zeeman Hamiltonian. It is well known from studies of quantum wells that this term leads to the additional heavy-hole light-hole mixing \cite{semenov_PRB_2003,koudinov_PRB_2006}, however in our case such additional mixing strength is more than order of magnitude weaker than mechanism presented in Eq. \ref{eq:mixing}  The second part depends on the third power of the momentum operators and therefore will be referred to as the cubic term of the LK Hamiltonian. 

The two terms of the LK lead to qualitatively different contributions to the hole g-factor.  Due to the form of the valence band mixing (Eq. (\ref{eq:mixing})), the Zeeman term of LK Hamiltonian results in fully anisotropic hole g-factor \cite{Koudinov_2004_PRB}. On the other hand, the presence of qubic term leads to isotropic contribution to the in-plane hole g-factor. In order to simplify the notation and isolate mechanisms of isotropic behavior of the in-plane hole g-factor we introduce an effective hole isotropy parameter
$\veps$ defined as:
\beq
\veps=\frac{\sqrt{3}q}{\sqrt{3}q+\lambda(4 \kp + 7 q)}.
\label{eq:espilon}
\eeq
With the use of this parameter, we obtain the expression for the in-plane hole g-factor tensor
\begin{equation}
\hat{g}_h=g_0 \frac{3q}{2\veps}
\begin{pmatrix}
\veps + (1-\veps)\cos 2\theta &-(1-\veps)\sin 2\theta \\[10pt]
(1-\veps)\sin 2\theta & -\veps + (1-\veps)\cos 2\theta
\end{pmatrix}, 
\label{eq:g_czynnik_dziury_eps}
\end{equation} 
defined by equation $\hat{H}^h_B=\mub \boldsymbol \sigma
\hat{g}_h\vecb{B}/2$, where $\boldsymbol \sigma$ are the Pauli matrices operating in the two-dimensional subspace of $\ket{\phi^{\pm}_{H}}$ lowest-energy hole states. It can be easily shown that $\veps=1$ corresponds to isotropic in-plane hole g-factor equal to $3g_0q/2$. On the other hand, $\veps=0$ results in entirely anisotropic character of the hole g-factor.     

Figure \ref{fig:example_map_and_fit}(b) presents an example numerical simulation of X$_d$ intensity for various directions of transverse magnetic field and polarization detection. The parameter $\veps$ was assumed 0.38 to best reproduce experimental data from Fig. \ref{fig:example_map_and_fit}(a). 

When the cubic term of LK Hamiltonian is neglected ($q=\veps=0$), the orientation of linear polarization of the dark exciton line $\gamma$ remains fixed for any direction of in-plane magnetic field. Its absolute orientation $\gamma=\theta$ depends only on the intrinsic anisotropy $\theta$ of the valence band mixing in a particular QD. On the other hand, in the absence of the Zeeman term ($\veps=1$) the hole g-factor exhibits isotropic behavior and the orientation of the dark exciton polarization $\gamma$ is determined by the direction of the magnetic field $\phii$ according to $\gamma = -\phii$. It should be stressed that these two orientations do not coincide, as the polarization orientation rotates in an opposite direction than applied magnetic field. These observations are similar to the previous report concerning polarization properties of CdSe/ZnSe QDs ensemble\cite{Kiessling_PRB_2006}  In the general case the dark exciton polarization direction $\gamma$ is governed by both terms in LK Hamiltonian and its dependence on the magnetic field direction is given by 
\beq
2\gamma=\operatorname{atan} \left[(1-2 \veps) \tan{(\phii+\theta)} \right] -\phii+\theta.
\label{eq:gamma}
\eeq 
This formula is obtained by analytical solution of the excitonic Hamiltonian $\hat{H}_{\mathrm{X}}$ (given in the Appendix) under an assumption that the value of anisotropic electron-hole exchange energy $\delta_1$ is much smaller than the splitting $\delta_0$ between dark  and bright levels \cite{footnote1}.

\subsection{Experimental data analysis and discussion}
~\\[-0.7cm]

\begin{figure}
\begin{center}
\includegraphics[width=85.9mm]{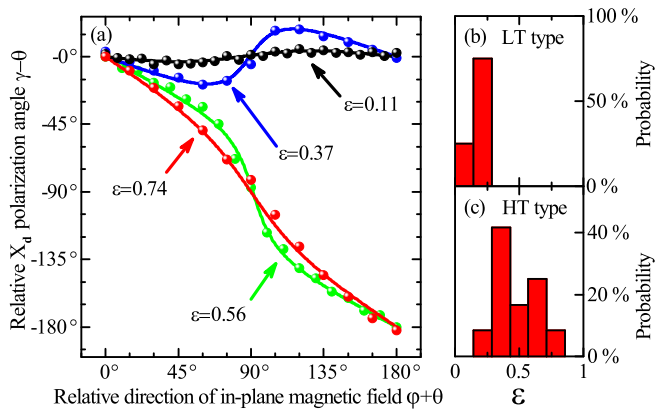}
\end{center}
\caption[]{(Color online) The orientation of linearly polarized light emitted from the recombination of the dark exciton as a function of direction of in-plane magnetic field. Solid lines indicate results of fitting Eq. (\ref{eq:gamma}) to obtained data (dots). Almost totally anisotropic nature of hole wave-function corresponds to low effective hole isotropy parameter $\veps=0.11$. In contrast, the QD characterized by $\veps=0.74$ demonstrates highly isotropic behavior of the hole wave-function. Intermediate cases are characterized with moderate effective hole isotropy parameters. (b), (c) Distribution of $\veps$ parameter among LT- and HT-type samples.}
\label{fig:gamma_epsilon_gallery}
\end{figure}

By fitting the analytical expression for $\gamma(\phii)$ to the experimental data we can efficiently determine the isotropy parameter $\veps$ for a given QD. A result of such a fit is presented in Fig. \ref{fig:example_map_and_fit}(c) as a solid line. Fig. \ref{fig:gamma_epsilon_gallery} shows the data and fits for a few different dots, demonstrating the scale of possible variation of the isotropy parameter among different QDs.

The simplicity of the presented procedure allowed us to study significant number of dots, grown by both LT and HT method. 
In general, QDs from LT-type sample  are characterized by low mean effective hole isotropy parameter $\bar{\veps}_{LT}=0.16 \pm 0.04$ (see Fig. \ref{fig:gamma_epsilon_gallery} (b)) Therefore, the hole wave function in LT QDs is highly anisotropic, which results in the almost constant polarization direction of the dark exciton emission line independent of the orientation of the in-plane magnetic field. Such QDs were commonly reported in literature \cite{Kowalik_2007_PRB, Smolenski_2012_PRB, Koudinov_2004_PRB, Leger_2007_PRB} and correspond to a negligible contribution of cubic term in LK Hamiltonian. Conversely, the QDs formed at higher temperature (HT-type) reveal a broad range of effective hole isotropy parameter, which spans over almost all values of $\veps$ from 0~to~1. As a consequence, the HT-type QDs are characterized by a higher mean value as well as higher standard deviation of effective isotropy parameter $\bar{\veps}_{HT}=0.46\pm 0.15$ (see Fig. \ref{fig:gamma_epsilon_gallery}(c)).

\begin{figure}
\begin{center}
\includegraphics[width=85.9mm]{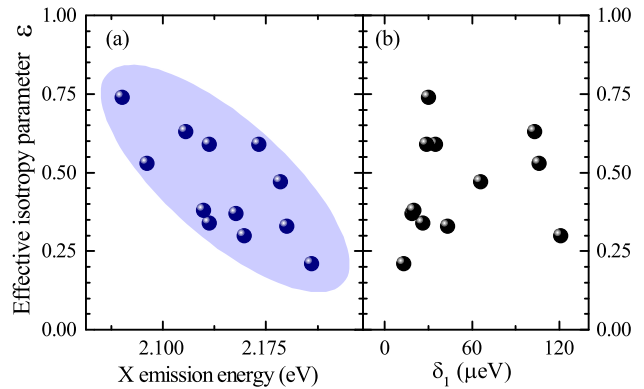}
\end{center}
\caption[]{(Color online) (a) Correlation of effective hole isotropy parameter with emission energy of dark exciton for HT sample. Ellipsoid in background is added to guide the eye. Lower energy QDs tend to have higher $\veps$ values. (b) There is no significant correlation between hole anisotropy and hole-electron exchange anisotropy $\delta_1$.  }
\label{fig:statistics_and_trend}
\end{figure}

Interestingly, we observe a significant correlation between the effective hole isotropy parameter and the QD emission energy among the HT dots. The larger effective hole isotropy parameter is observed for QDs with lower emission energies of the neutral exciton (Fig. \ref{fig:statistics_and_trend}(a)). However, there is no correlation between $\veps$ and electron-hole exchange anisotropy $\delta_1$ (Fig. \ref{fig:statistics_and_trend}(b)). These observations can be explained in terms of the average anisotropy of the QD sampled by hole wave function. Lower emission energy is related to a larger extension of the hole wave function, which then can average out local anisotropy of different parts of the confining potential. Analogically, higher emission energy is related to a smaller wave function, and therefore even a modest local anisotropy of a QD can strongly influence the hole wave function resulting in its high anisotropy and small $\veps$ value.  

The observed difference of standard deviation of $\veps$ between LT- and HT-type samples can be related to easier diffusion of interface composition inhomogeneities at higher temperatures, which results in more isotropic hole wave function for HT-type samples. 

\section{Summary}
We have presented a simple method of determining anisotropy properties of the hole wave function in self-assembled QDs using the measurement of the dark exciton PL. The in-plane magnetic field makes the dark exciton optically active. Moreover, the linear polarization orientation of the dark exciton emission line strongly depends on the direction of the applied magnetic field. The observed changes of the polarization direction are well described by the presented model and allows the determination of effective hole wave function anisotropy as well as the in-plane hole g-factor anisotropy. We show experimental results for the two types of self-assembled QDs, which differ in growth temperature. The QDs formed at higher temperatures reveal much more isotropic character of the hole wave function than those grown at lower temperatures. We also observed CdTe QD with almost completely isotropic hole wave function, which indicates that the cubic term in LK Hamiltonian might have leading contribution to the hole g-factor. The presented method of determining anisotropy properties of the hole wave function can be applied to any system of self assembled QDs.       

\section{Acknowledgements}
We thank Witold Bardyszewski for fruitful discussions. This work was supported by the Polish Ministry of Science and Higher Education in years
2012$-$2017 as research grant "Diamentowy Grant"; by the Polish National Science Centre under decisions DEC-2011/02/A/ST3/00131, DEC-2012/05/N/ST3/03209, DEC-2012/07/N/ST3/03130; and by the Foundation for Polish Science (FNP) through "Mistrz" subsidy. One of us (T.S.) was supported the Foundation for Polish Science (FNP) through START programme. The project was carried out with the use of CePT, CeZaMat, and NLTK infrastructures financed by the European Union - the European Regional Development Fund within the Operational Programme ''Innovative economy'' for 2007 - 2013. 

\appendix
\section{Hamiltonian and analytical approximation}
\label{app:matrix}

We used full Hamiltonian of a neutral exciton in self-assembled QD equal to 
\begin{equation} \label{eq:Hamiltonian_pelen_ekscytonu_nowa_baza}
\hat{H}_{\mathrm{X}}=\sfr{2}
\begin{pmatrix}
-\delta_0				&		B\!\cdot\!M_e		&		B\!\cdot\!M_h		&	\delta_2 			\\[5pt]
B\!\cdot\!M_e^\dag 		&	\delta_0			&	\delta_1			&	B\!\cdot\!M_h	\\[5pt]
B\!\cdot\!M_h^\dag 		&	\delta_1			&	\delta_0			&	B\!\cdot\!M_e	\\[5pt]
\delta_2						&	B\!\cdot\! M_h^\dag 	&	B\!\cdot\!M_e^\dag 	&	-\delta_0
\end{pmatrix},
\end{equation}
written in the basis $\ket{\ua_e,\phi^+_H},\ \ket{\da_e,\phi^+_H},\ \ket{\ua_e,\phi^-_H},\ \ket{\da_e,\phi^-_H}$, where $M_e=\mu_B g_e \exp(-\ii \phii)$, and
\begin{equation}
M_h=\frac{\mu_B3g_0q}{2\veps} \left[ \veps e^{\ii \phii} + (1-\veps) e^{-\ii( 2 \theta +\phii)} \right],
\end{equation}
with $\veps$ defined by Eq. (\ref{eq:espilon}). Under the assumption that $\delta_2\ll \delta_1 \ll \delta_0$ the above Hamiltonian can be diagonalized analytically. The energies of mostly dark eigenstates $|\psi_{\pm}\rangle$ are given by
\begin{equation}
E_\pm=-\sfr{2} \sqrt{ \delta_0^2 + B^2 ( |M_e|\pm |M_h| )^2 }.
\end{equation} 
The optical properties of both of these states are determined by their overlap with anti-parallel spin states of the neutral exciton, which yield $\eta_\pm=\bra{\da_e,\phi^+_H}\psi_{\pm}\keta{}$ and $\xi_\pm=\bra{\ua_e,\phi^-_H}\psi_{\pm}\keta{}$. For each of the dark exciton states, the magnitudes of both of these terms are equal $|\eta_\pm|=|\xi_\pm|$. With the use of these parameters and under the assumption that the amplitude of the valence-band mixing $\lambda$ is small\cite{Koudinov_2004_PRB,Leger_2007_PRB} one can obtain the expression for oscillator strengths of $|\psi_{\pm}\rangle$-related transitions at linear detection angle $\beta$ in the presence of the in-plane magnetic field applied at  direction~$\phii$:
\begin{equation}
f_\pm=\sfr{2}\frac{B^2 M_{\pm}^2}{(\delta_0-2E_\pm)^2+B^2 M_{\pm}^2}\left[1\pm\cos(2\beta-2\gamma)\right],
\label{eq:proste_rownanie_na_f}
\end{equation}
where $\gamma$ is given by Eq. (\ref{eq:gamma}) and $M_{\pm}=|M_e|\pm|M_h|$. As such, the optical transitions from both of the mostly dark states $|\psi_\pm\rangle$ are fully linearly polarized in perpendicular directions defined by $\gamma$ and $\gamma+90^\circ$. However, when $|M_e|$ has similar value to $|M_h|$, $|\psi_+\rangle$ has the dominant contribution to the emission intensity. Given that both of the $|\psi_\pm\rangle$ states are almost degenerate at low magnetic field used in our experiments, they appear in the PL spectrum as a single line, which is partially linearly polarized at $\gamma$ direction.
\bibliographystyle{apsrev_my}
\bibliography{anizo_voigt_dark_biblio}

\end{document}